\documentclass[12pt,preprint]{aastex}

\newcommand\lsim{\mathrel{\spose{\lower 3pt\hbox{$\mathchar"218$}}
     \raise 2.0pt\hbox{$\mathchar"13C$}}}
\newcommand\gsim{\mathrel{\spose{\lower 3pt\hbox{$\mathchar"218$}}
     \raise 2.0pt\hbox{$\mathchar"13E$}}}

\newcommand{\chandra}{{\sl Chandra\/}}

\newcommand{\vsgr}{V1223~Sgr}
\newcommand{\exhya}{EX~Hya}
\newcommand{\sscyg}{SS~Cyg}
\newcommand{\ugem}{U~Gem}
\newcommand{\vaql}{V603~Aql}
\newcommand{\aopsc}{AO~Psc}
\newcommand{\gkper}{GK~Per}
\newcommand{\ngc}{NGC~1068}

\shorttitle{Two Types of X-ray Spectra in CVs}
\shortauthors{Mukai et al.}

\begin{document}


\title{Two Types of X-ray Spectra in Cataclysmic Variables}


\author{K. Mukai\altaffilmark{1,2}}
\affil{Code 662, NASA/Goddard Space Flight Center, Greenbelt, MD 20771, USA.}
\email{mukai@milkyway.gsfc.nasa.gov}

\and

\author{A. Kinkhabwala, J.R. Peterson, S.M. Kahn, F. Paerels, }
\affil{Columbia Astrophysics Laboratory, Columbia University,
	550 West 120th Street, New York, New York 10027, USA.}
\email{ali@astro.columbia.edu,jrpeters@astro.columbia.edu,
	skahn@astro.columbia.edu,frits@astro.columbia.edu}


\altaffiltext{1}{Also Universities Space Research Association}
\altaffiltext{2}{Also Columbia Astrophysics Laboratory, Columbia University,
	550 West 120th Street, New York, New York 10027, USA}


\begin{abstract}

We present \chandra\ HETG spectra of seven cataclysmic variables.
We find that they divide unambiguously into two distinct types.
Spectra of the first type are remarkably well fit by a simple cooling
flow model, which assumes only steady-state isobaric radiative cooling.
The maximum temperature, $kT_{\mathrm{max}}$, and the normalization,
which provides a highly precise measurement of the accretion rate, are
the only free parameters of this model.  Spectra of the second type
are grossly inconsistent with a cooling flow model.  They instead exhibit
a hard continuum, and show strong H-like and He-like ion emission but
little Fe~L-shell emission, which is consistent with expectations for
line emission from a photoionized plasma.  Using a simple photoionization
model, we argue that the observed line emission for these sources
can be driven entirely by the hard continuum.  The physical significance
of these two distinct types of X-ray spectra is also explored.

\end{abstract}




\section{Introduction}

Cataclysmic variables (CVs) are interacting binaries in which the
accreting object (the primary) is a white dwarf (see \citealt{tome}
for a review).  X-ray emission in CVs is most likely associated with
the accretion process, which is capable of shock-heating accreted 
material up to high temperatures ($kT_{\mathrm{max}}\sim$ 10--50~keV).
Their accretion geometry is strongly influenced by the
magnetic field of the primary.  Non-magnetic systems have undisrupted
accretion disks that connect to the white dwarf surface via a
boundary layer \citep{PR1985}.  In magnetic systems,
the accretion stream follows the primary's magnetic field lines,
and is close to vertical when it
hits the surface (see, e.g., \citealt{A1973}).  In both cases,
the emergent X-ray spectrum is expected to be
the sum of emission from plasmas over a continuous temperature distribution, 
from the shock temperature to the white dwarf photospheric temperature.  

\section{Observations and Spectra}

Seven CVs were observed with \chandra\ HETG by 2002 March;
we have obtained these data from the archive and analyzed them
using CIAO 2.2.1.  The objects are four magnetic CVs of the intermediate
polar (IP) class (\exhya, \vsgr, \aopsc, and \gkper), two dwarf novae
(\sscyg\ and \ugem), and one old nova (\vaql).  For CVs with multiple HETG
observations, we have chosen to analyze one spectrum per system
(quiescent spectrum of \sscyg, and one of two similar spectra of
\gkper).  Previous publications from these data exist for
\exhya\ \citep{Mea2001,M2002} and \ugem\ \citep{Sea2002}.  

These seven X-ray spectra divide unambiguously into two types.  Spectra
of the first type are well fit by a simple cooling flow model, whereas
spectra of the second type are well described by a simple model of a 
photoionized plasma.  Combined first-order ($m=\pm1$) MEG spectra
of cooling flow CVs and of photoionized CVs are shown in 
Figs.~\ref{fig:cool} and \ref{fig:phot}, respectively, and are discussed
below in \S\ref{sec:cool} and \S\ref{sec:phot}, respectively.

\section{Spectral Analysis}

\subsection{Cooling Flow Spectra}\label{sec:cool}

\begin{figure}
\plotone{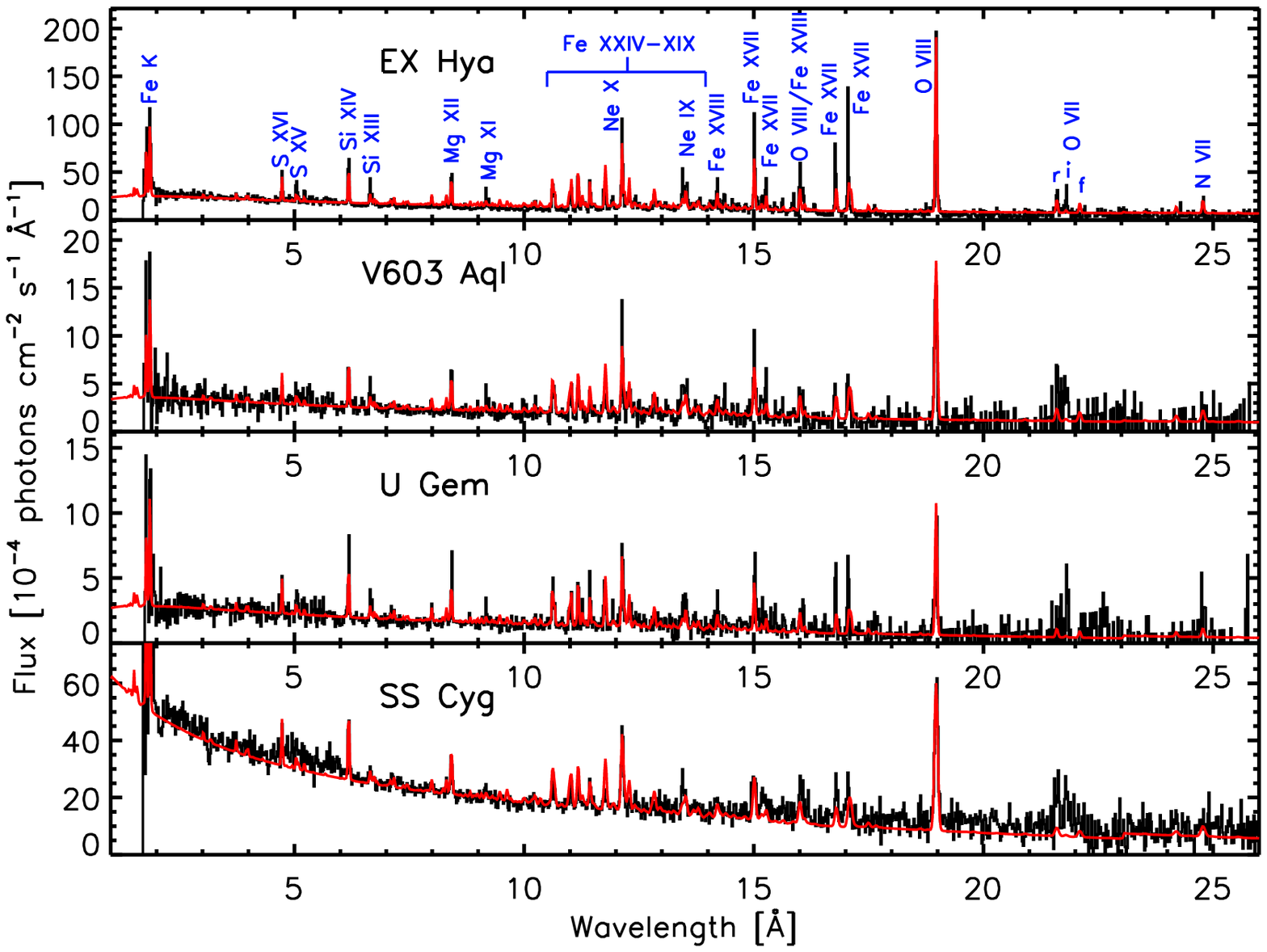}
\caption{\chandra\ HETG spectra (MEG, $m=\pm1$ orders) of 4 CVs that exhibit
cooling flow spectra.  Data are shown in black with the corresponding
cooling flow model shown in red.}
\label{fig:cool}
\end{figure}

The four CV spectra of \exhya, \vaql, \ugem, and \sscyg\
shown in Fig.~\ref{fig:cool} are strikingly similar.  
These spectra exhibit a smooth continuum, strong H- and He-like ion
emission lines of O, Ne, Mg, Al, Si, and S, and strong emission from
ions spanning the entire Fe~L-shell complex (\ion{Fe}{17}-\ion{Fe}{24}).
In the HEG spectra (not shown), H- and He-like Fe lines are strong,
while the intensity of the Fe fluorescence line varies from system to system.

The observed wide range and strengths of emission from H- and He-like ions, 
the level and shape of the continuum (suggesting bremsstrahlung), and, in 
particular, the comparable strengths of emission across the full range of 
Fe~L-shell ions are all consistent with expectations for a multitemperature 
thermal plasma with a relatively flat emission measure distribution.
The flatness of this distribution is further indicative 
of an isobaric cooling flow, which assumes only that the gas releases
all of its energy in the form of optically-thin radiation as it cools in 
a steady-state flow.  The two main parameters of this model are the 
maximum temperature, $kT_{\mathrm{max}}$, and the overall 
normalization, which directly gives the total mass flow rate.
Fits to the spectra using {\tt mkcflow} \citep{MS1988}
(with a uniform velocity broadening and an interstellar absorber)
in {\tt xspec} \citep{A1996} are shown in red in Fig.~\ref{fig:cool} with
model parameters listed in Table~\ref{tab:cool}.  We obtain the
same $kT_{\mathrm{max}}$ of 20~keV for \exhya, \vaql, and \ugem, but 
a $kT_{\mathrm{max}}$ of 80~keV for \sscyg, accounting for the stronger
continuum in this source.  The total mass flow rates
listed in Table~\ref{tab:cool}, which we have not scaled to the
best-estimate distances to these CVs, are nonetheless entirely reasonable.

These high resolution observations unambiguously demonstrate
that post-shock plasmas in some CVs are cooling flows, as originally
pointed out by \cite{FN1977} and applied to CVs by \cite{Dea1995}.
Interestingly, the {\tt mkcflow} model developed for clusters of galaxies
can be used to fit CV spectra, while it is now known that clusters are
poorly described by such a model \citep{Pea2001,Pea2002}.

Our models (Fig.~\ref{fig:cool}) generally predict line strengths to within
a factor of two, setting the maximum level for abundance differences
from our assumption (solar).  Some of these discrepancies, such as in
the \ion{Fe}{17} line complex, are more likely due to effects arising
from line opacity, density, or UV irradiation, rather than abundance
(especially given the concordance of the \ion{Fe}{18}--\ion{Fe}{24} lines).
We also note that He-like triplets (e.g., \ion{O}{7}, \ion{Ne}{9}, and
\ion{Mg}{11}), whenever data quality is sufficiently high to be resolvable,
show evidence for forbidden-to-intercombination line conversion through
excitation of the long-lived 2 $^3$S$_1$ level up to the 2 $^3$P multiplet
\citep{GJ1969}; both electron collisional excitation (e.g., \citealt{PD2000})
and photoexcitation due to UV photons from the white dwarf surface
(e.g., \citealt{Kea2001}) are potentially important in driving this conversion.

\subsection{Photoionized Spectra}\label{sec:phot}

\begin{figure}
\plotone{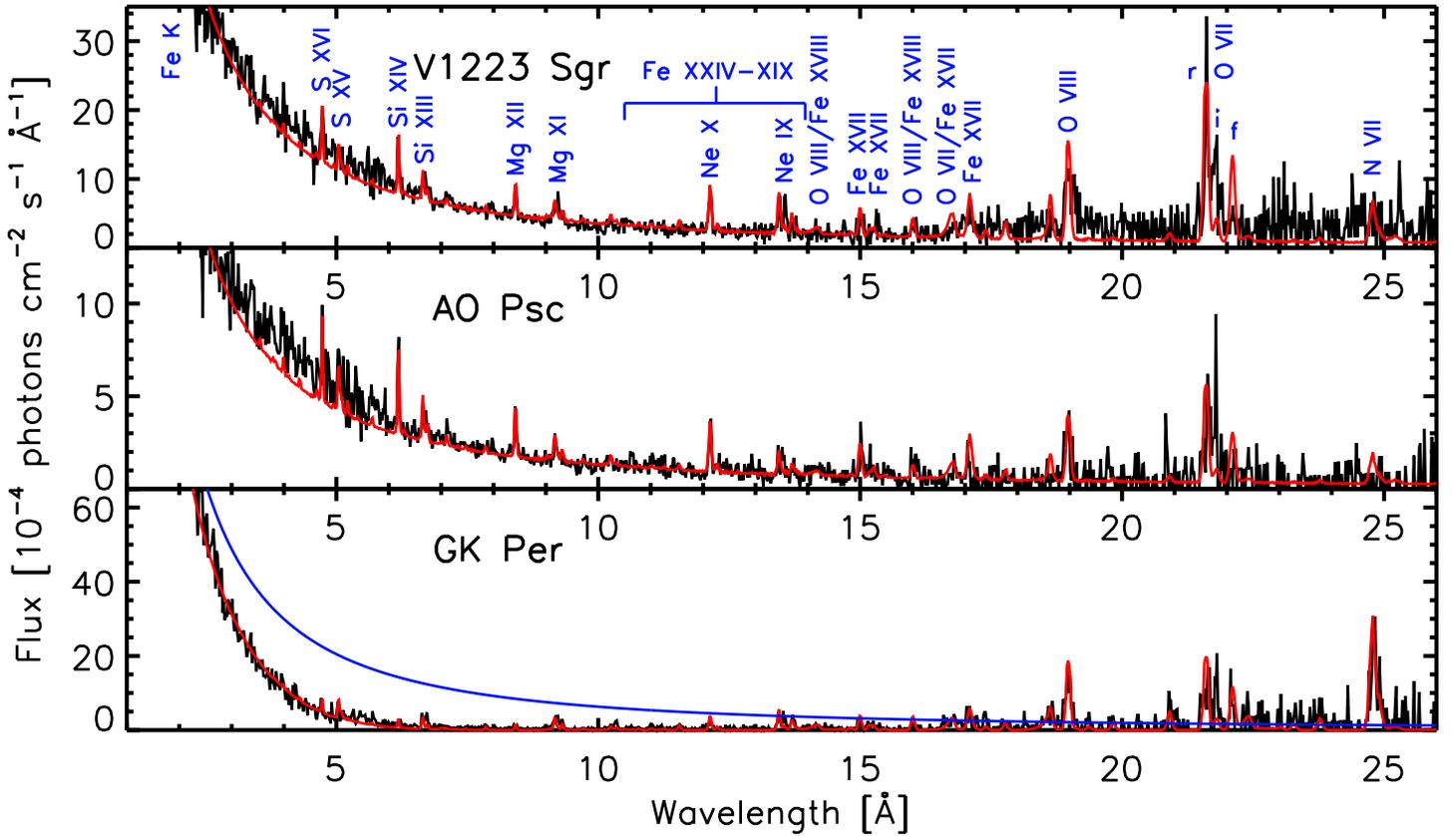}
\caption{\chandra\ HETG spectra (MEG, $m=\pm1$ orders) of 3 CVs that exhibit
photoionized models.  Data are shown in black with the corresponding
photoionization model shown in red.  The blue line in the bottom panel
shows the inferred intrinsic continuum for \gkper.}
\label{fig:phot}
\end{figure}

The spectra shown in Fig.~\ref{fig:phot} are noticeably different from
those shown in Fig.~\ref{fig:cool}, but are very similar to one another.
In particular, \vsgr\ and \aopsc\ both have a hard, power-law-like
continuum, and strong lines from medium Z elements, but the Fe L-shell
emission is limited only to \ion{Fe}{17} lines, with no detectable
\ion{Fe}{18}-\ion{Fe}{24} lines.  This does not indicate a low Fe abundance,
since we detect strong K$\alpha$ lines (H-like, He-like, and fluorescence)
in the HEG spectra (not shown).  \gkper\ has a continuum that is harder still,
and shows H- and He-like ions but little Fe L-shell emission.  This type of
emission line spectrum is incompatible with cooling flow plasma.  Instead,
it resembles the expectations for photoionized plasma \citep{L1999,Sea2000},
and the observed spectrum of the prototype Seyfert~2, \ngc, for which
photoionization origin has been demonstrated \citep{Kea2002}, confirming
the earlier suggestion by \cite{Kea1993}.

We model these spectra with the {\tt xspec} photoionization model 
{\tt photoion}\footnote{http://xmm.astro.columbia.edu/photoion/photoion.html} 
\citep{Kea2002,Kea2003}.  We assume that the observed continuum of
\vsgr\ and \aopsc\ is that which drives the line emission, while a similar
continuum (shown as the blue line in Fig.~\ref{fig:phot}) drives the
photoionization in \gkper\ but is obscured along our particular line of sight.
We also assume the line emission is unabsorbed in all three systems.
Furthermore, we assume that all lines remain
completely unsaturated at all ionic column densities, due to radial velocity
distribution widths of thousands of km~s$^{-1}$.  This last assumption is
necessary to explain the strong resonance lines in the He-like triplets,
which are not predicted by traditional photoionization models \citep{PD2000}.
With these assumptions, we find that a simple model of a photoionized 
cone of plasma fits the data very well.  We show the model fits in
Fig.~\ref{fig:phot} in red (showing the absorbed continuum plus unabsorbed
lines in the case of \gkper), and list the corresponding model parameters
in Table~\ref{tab:phot}.  Included in this table are the lower limits to
the accretion rates necessary to power the photoionizing continuum.
These values are similar to the mass flow rates in the cooling-flow CVs.


The presence of line emission from \ion{O}{7} to \ion{S}{16} in all
spectra is consistent with a broad, relatively flat 
distribution in log ionization parameter, as inferred for
\ngc\ \citep{Kea2002}, and with near solar abundances.  The exception is
the large nitrogen to oxygen ratio in \gkper, mirroring the unusual
UV line ratios \citep{Mea1997}, and may suggest an abundance anomaly
for this system with an evolved secondary \citep{K1964}.
As with the cooling flow cases, the He-like triplets again exhibit conversion
of the forbidden to intercombination lines, suggesting high electron density 
and/or a strong, ambient UV radiation field.

\section{Discussion}

We have shown that a simple cooling flow model is capable of reproducing
the X-ray spectra of four CVs, while a photoionization model can successfully
reproduce the strikingly different spectra of three other CVs.  The latter
are all magnetic systems of the IP type, while the former is a mixture
of \exhya, an unusual IP, and non-magnetic systems.  Both the fact that
the X-ray spectral types almost coincide with the classification into magnetic
and non-magnetic systems, and the somewhat surprising fact that this match
is not exact, requires explanation.

The difference between the two X-ray spectral types may be the
specific accretion rate (accretion rate per unit area).
In most IPs, the magnetic field collimates the flow onto small
regions ($<$0.2\% of the white dwarf surface in XY~Ari: \citealt{H1997}).
\exhya, however, is thought to have a much lower specific accretion rate
than typical IPs, creating a tall, lower density, shock \citep{Aea1998}.
The resultant top-hat geometry of the post-shock region will allow the
cooling flow X-rays to escape the post-shock region from the side without
having to travel through the pre-shock flow.  The boundary layer in
non-magnetic CVs probably cover a much larger area than the spot on XY~Ari.
This would lead to a specific accretion rate that is always low enough to
allow the X-rays to escape freely.  In this view, all CVs with low specific
accretion rate (all non-magnetic systems, as well as some magnetic systems)
should show a cooling flow-like X-ray spectrum.

On the other hand, typical IPs such as \vsgr\ and \aopsc\ probably have
a high specific accretion rate, leading to a pillbox geometry which does
not allow the cooling flow spectrum to escape freely.  If true, these
systems are also powered by the cooling flow continuum (compare the measured
$\dot{M}$ in Table~\ref{tab:cool} with the inferred values in
Table~\ref{tab:phot}), which is differentially absorbed.  The apparent
lack of the cooling-flow line emission may be due to a larger optical depth
along the line of sight to the post-shock region than to the photoionized
plasma, allowing the latter to dominate the observed spectrum.  Although
a model of a multitemperature bremsstrahlung continuum with multiple partial
covering absorbers can reproduce the observed continuum shape, this does beg
the question of why we observe such highly similar hard continua in \vsgr\ and
\aopsc.   If this cannot be explained without an excessive degree of
fine-tuning, we may be forced to explore an alternative origin of the ionizing
continuum.

Evidence that the photoionized emission arises from pre-shock flow
comes from the inferred large radial velocity distribution,
which is present only in the near free-fall pre-shock flow.
Further evidence may come from \gkper.  The shorter-wavelength, 
higher-ionization line strengths are significantly weaker in this source
than in the other two.  If, for \gkper, we discard the assumption that
the line emission is unabsorbed, and instead assume that its line emission
spectrum is intrinsically similar to that of \vsgr\ and \aopsc, then
\gkper\ must be differentially obscured, with obscuration increasing
with ionization parameter all the way down to the 
photoionizing continuum.  This favors a spatially-stratified ionization model, 
which should arise naturally in an accretion flow, rather than the intrinsic
density distribution that leads to the range of ionization in
\ngc\ \citep{Bea2002}.

The existence of two types of X-ray spectra in the HETG range among
accreting CVs is beyond doubt.  However, our interpretation is admittedly
somewhat tentative, and that other system parameters (such as the primary
mass and the inclination angle) may also play an important role in shaping
the observed X-ray spectra.  It is also possible that there are yet other types
of CV X-ray spectrum in the HETG range, such as the wind-scattered X-rays
seen at longer wavelengths (see, e.g., \citealt{Kuea2002}).    Further high
resolution X-ray spectroscopy of a larger sample of CVs, and more detailed
analyses of the existing spectra, should be helpful in clarifying both points.

\acknowledgments

We thank Ming Feng Gu for previous extensive help with his atomic code FAC,
and Coel Hellier and Chris Mauche for useful discussion.
The Columbia University team is supported by NASA.  AK acknowledges 
additional support from a NASA GSRP fellowship.

\clearpage


\clearpage

\begin{deluxetable}{lcccc}
\tablecaption{Cooling Flow CVs.}
\tablecolumns{5}
\tablewidth{3.33in}
\tablehead{\colhead{Model} & \colhead{\exhya} & \colhead{\vaql} & \colhead{\ugem} & \colhead{\sscyg}\\
\colhead{Parameters} & \colhead{1706\tablenotemark{a}} & \colhead{1901\tablenotemark{a}} & \colhead{647\tablenotemark{a}} & \colhead{3454\tablenotemark{a}}}
\startdata
$N_{\mathrm{H}}$ [cm$^{-2}$]\  	& 2e20     &  2e20   &  1e21   &  7e20   \\
$\sigma_v$ [km s$^{-1}$]\tablenotemark{b} & 200     &  400   & 300    & 550    \\
$kT_{\mathrm{max}}$ [keV] & 20     &  20   & 20    & 80    \\
$\dot{M}\,d_{100}^{-2}$\ \tablenotemark{c}  & 3.9e-11 & 5.6e-12 & 4.4e-12 & 2.6e-11 \\
\enddata
\tablenotetext{a}{ObsID.}
\tablenotetext{b}{Observed line width, assuming a constant velocity broadening
	for all lines.}
\tablenotetext{c}{Mass flow rate for an assumed distance of 100 pc
	in $\mathrm{M}_{\odot}$~yr$^{-1}$.}
\label{tab:cool}
\end{deluxetable}

\begin{deluxetable}{lccc}
\tablecaption{Photoionized CVs.}
\tablecolumns{4}
\tablewidth{3.33in}
\tablehead{\colhead{Model} & \colhead{\vsgr} & \colhead{\aopsc} & \colhead{\gkper}\\
\colhead{Parameters} & \colhead{649\tablenotemark{a}} & \colhead{1898\tablenotemark{a}} & \colhead{3454\tablenotemark{a}}}
\startdata
$f$\ \tablenotemark{b}   & 0.5     & 0.5    & 0.5    \\
$N^{\mathrm{intrinsic}}_{\mathrm{H}}$ [cm$^{-2}$]\ \tablenotemark{c}   	&  0    &   0  & 7e22    \\
$\sigma_v$ [km s$^{-1}$]\tablenotemark{d} & 600     &  600   & 600  \\
$L_X\,d_{100}^{-2}$\ \tablenotemark{e}   & 1.6e32     & 6e31    & 2.9e32    \\
$\dot{M}\,d_{100}^{-2}$\ \tablenotemark{f}  & $>$9.4e-12 & $>$3.5e-12 & $>$1.7e-11 \\
\hline
\multicolumn{4}{c}{Ionic Column Densities [cm$^{-2}$]} \\
\hline
\ion{N}{6} 	& 8e17     & 5e17    & 2e18    \\
\ion{O}{7} 	& 2.5e18   & 1.3e18  & 1.2e18  \\
\ion{O}{8} 	& 1.7e18   & 1e18    & 1.2e18  \\
\ion{Ne}{9}	& 4e17     & 2e17    & 2e17    \\
\ion{Ne}{10}	& 5e17     & 5e17    & 1.5e17  \\
\ion{Mg}{11}	& 1.7e17   & 2e17    & 1e17    \\
\ion{Mg}{12}	& 3.2e17   & 4.7e17  & 5e16    \\
\ion{Si}{13}	& 1.7e17   & 2.5e17  & 7e16    \\
\ion{Si}{14}	& 5e17     & 7e17    & 7e16    \\
\ion{S}{15}	& 1.5e17   & 2.5e17  & 1e17    \\
\ion{S}{16}	& 4e17     & 6e17    & 1e17    \\
\ion{Fe}{17}	& 2e17     & 2e17    & 1e17    \\
\enddata
\tablenotetext{a}{ObsID.}
\tablenotetext{b}{Covering fraction $f=\Omega/4\pi$ for cone geometry.}
\tablenotetext{c}{Required obscuration of inferred continuum for \gkper.}
\tablenotetext{d}{Observed line width, assuming a constant velocity broadening
	for all lines.}
\tablenotetext{e}{Bolometric power-law $L(E)\propto E^{-0.3}$ 
(13.6~eV$<E<10$~keV) luminosity for an assumed distance of 100 pc
	in ergs~s$^{-1}$.}
\tablenotetext{f}{Minimum accretion rate for an assumed distance of 100 pc
	in $\mathrm{M}_{\odot}$~yr$^{-1}$.}
\label{tab:phot}
\end{deluxetable}

\end{document}